%% file: stau_susy02_rev.tex
\documentclass[12pt,fleqn]{article}

\usepackage{amssymb,graphicx,epsfig}
\usepackage{lscape,graphics,amsmath,geometry}
\usepackage{latexsym}
\usepackage[dvips]{color}
\newcommand {\beq} {\begin{eqnarray}}
\newcommand {\eeq} {\end{eqnarray}}
\def\lsim{\raise0.3ex\hbox{$\;<$\kern-0.75em\raise-1.1ex\hbox{$\sim\;$}}}
\def\gsim{\raise0.3ex\hbox{$\;>$\kern-0.75em\raise-1.1ex\hbox{$\sim\;$}}}
\def\greaterthansquiggle{\raise.3ex\hbox{$>$\kern-.75em\lower1ex\hbox{$\sim$}}}
\def\lessthansquiggle{\raise.3ex\hbox{$<$\kern-.75em\lower1ex\hbox{$\sim$}}}

\newcommand{\la}{\label}

\newcommand{\ci}{\cite}
\newcommand{\beqn}{\begin{eqnarray}}
\newcommand{\eeqn}{\end{eqnarray}}
\newcommand{\bequ}{\begin{equation}}
\newcommand{\eequ}{\end{equation}}
\newcommand{\bsl}{\begin{sloppypar}}
\newcommand{\esl}{\end{sloppypar}}
\newcommand{\bm}{\boldmath}
\newcommand{\um}{\unboldmath}

\begin{document} 
\hfill hep-ph/0211040\\
\hfill .\\
\vskip .4cm

\begin{center}
{\Large\bf
Impact of Tau Polarization for the\\[.3em] 
determination of high \bm$\tan\beta$ \um 
 and \bm $A_{\tau}$\um \footnote[4]{Talks given at the 10th International
Conference on Supersymmetry and Unification of Fundamental Interactions,
SUSY02, DESY, Hamburg 2002}}
\vskip 1.5em

{\large
{
\underline{E. Boos}$^{a,b}$\footnote[1]{Speaker
at SUSY02, boos@theory.sinp.msu.ru}, 
\underline{G.
Moortgat-Pick}$^{b,c}$\footnote[7]{Speaker at SUSY02, 
gudrid@mail.desy.de}, H.U. Martyn$^{d}$, 
M. Sachwitz$^{e}$, A.Vologdin$^{a}$
}                     
}\\[3ex]
\end{center}

{\footnotesize \it 
\noindent
$^{a}$ Skobeltsyn Institute of Nuclear Physics, Moscow State University,
119992 Moscow, Russia\\
$^{b}$ DESY, Deutsches Elektronen-Synchrotron, D-22603 Hamburg, Germany \\
$^{c}$ II. Institute for Theoret. Physics, University of Hamburg, D-22761
Hamburg, Germany \\
$^{d}$ Rheinisch-Westf\"alische Technische Hochschule, D-52074 Aachen, 
Germany \\
$^{e}$ DESY, Deutsches Elektronen-Synchrotron, D-15738  Zeuthen, Germany
}\\
\vskip .5em
\par

\begin{abstract}
In order to determine the fundamental MSSM parameters $M_1$, $M_2$,
$\mu$ and $\tan\beta$ the tau polarization from $\tilde{\tau}$ decays
can be explored as a `bridge' between the gaugino/higgsino and the
stau sector in particular in the high $\tan\beta$ range. 
Even in the case of high 
$\tan\beta$ an accuracy of $\delta(\tan\beta)\approx 5\%$ with an 
simultanous determination of $A_{\tau}$ is possible without 
assuming a specific SUSY breaking scheme.
\end{abstract}

\input{Introduction_fin}

\input{Stau_fin}

\input{Strategy_fin}

\input{Conclusions_fin}

\vspace{.2cm}
The work of E.B. and A.V. was partly supported by
the INTAS 00-0679, CERN-INTAS 99-377, and RFBR 01-02-16710 grants.
E.B. thanks the Humboldt Foundation for the Bessel Research Award
and DESY for the kind hospitality.

\end{document}

%% file: Introduction_fin.tex
\section{Introduction}
The Minimal Supersymmetric extension of the Standard Model (MSSM) is
one of the most promising extensions of the Standard Model (SM).
However, SUSY has to be broken and 
in the unconstrained version of the MSSM a
parameterization of all possible soft SUSY breaking terms 
leads to 105 new parameters in addition to the ones of the SM.

A linear collider (LC) is, due to its clear
signatures, the most promising tool 
for revealing the underlying structure of the physics beyond the SM
as e.g. the precise determination of these parameters.

\begin{sloppypar}
Strategies to determine the fundamental parameters in the
chargino/neutralino sector have already been worked out in
\ci{Parameters,CKMZ} and references therein. However, for
$\tan\beta>10$, the
gaugino/higgsino sector is weak dependent on $\tan\beta$ so that its
determination becomes rather inaccurate.  
We therefore concentrate in this paper on the production of
the $\tau$ SUSY partners $\tilde{\tau}_{1,2}$ and their decays into
$\tilde{\chi}^0_1$ and using the polarization of the $\tau$'s for the
accurate determination of high $\tan\beta$.
Since the $\tau$ polarization
involves simultanously mixing parameters from both neutralino and stau
sectors it plays the role of a `bridge' between these two sectors.
\end{sloppypar}

\begin{sloppypar}
The importance of the $\tau$ polarization has already been pointed out
in \ci{Noji,Roy}. We derive the compact formula for 
the polarization including the general mixing in the neutralino sector
 and show
in which regions of MSSM parameter space the polarization will be 
a suitable observable.
Contrary to the former studies we assume no specific character of the LSP 
$\tilde{\chi}^0_1$  or specific GUT relations between the underlying 
gaugino parameters. 
We show with an numerical example that
the polarization on $\tau$'s is well suited for a rather accurate
determination even of high 
$\tan\beta$ as well as for a simultanous determination of $A_{\tau}$.
\end{sloppypar}

%% file: Stau_fin.tex
\section{The Stau Sector}
\vspace{-.2cm}
\subsection{Masses and Mixing}
\vspace{-.2cm}
Since the $\tau$ lepton has the largest Yukawa
coupling of the three lepton families 
the weak eigenstates
$\tilde{\tau}_{L,R}$ mix to the mass eigenstates $\tilde{\tau}_{1,2}$,
where the mass matrix is given by:
\begin{eqnarray}
{\cal M}^2_{\tilde{\tau}} = 
      \left( \begin{array}{cc}
        M^2_L + m_{\tau}^2  + D_L & m_{\tau}(A_{\tau} -\mu \tan\beta)  \\
 m_{\tau}(A_{\tau} -\mu \tan\beta)&  
M^2_E + m_{\tau}^2  + D_R \end{array} \right)
=\left( \begin{array}{cc}
        m_{LL}^2 & m_{LR}^2 \\
 m_{LR}^2&  m_{RR}^2 \end{array} \right)
\label{eq_21_002}
\end{eqnarray}
with the D-terms 
$D_L = (-{\frac{1}{2}} + \sin^2 \theta_W ) \cos(2 \beta)  m^2_Z$ and
$D_R = - \sin^2 \theta_W \cos(2 \beta)  m^2_Z$,
the trilinear slepton--Higgs $\tilde{\ell}^{*}_R$-$\tilde{\ell}_L$-$H_1$
coupling $A_{\tau}$, the higgsino mass parameter $\mu$, the ratio
of the Higgs expectation values $\tan\beta=v_2/v_1$ 
 and the SU(2) doublet 
(singlet) mass parameters $M_L$ ($M_E$). The mass parameters 
$m_{LL}^2$, $m_{RR}^2$ have to be positiv for $\tan\beta>1$, whereas 
the sign of the off--diagonal terms 
$m_{LR}^2$ depends on $A_{\tau}$ and $\mu$.

The mass eigenvalues are given by
\begin{equation}
m_{\tilde{\tau}_{1,2}}^2=
\frac{1}{2}[m_{LL}^2+m_{RR}^2\mp
\sqrt{(m_{LL}^2-m_{RR}^2)^2+4(m_{LR}^2)^2}], \label{eq_massev}
\end{equation}
and a large mass difference may be quite natural.
In many SUSY scenarios
$\tilde{\tau}_1$ is similar light as light charginos/neutralinos.
A future high
${\cal L}$ LC will be well suited to measure the masses
with an high accuracy of e.g. about $\delta(m_{\tilde{\tau}_1})\sim 0.6$~GeV
\ci{TDR}.       

The stau mixing angle $\theta_{\tilde{\tau}}$, $[0,\pi]$ is given by:
\begin{equation}
\tan (2 \theta_{\tilde{\tau}})= \frac{- 2 m_{LR}^2}{m^2_{RR}-m^2_{LL}}=:\xi
\la{eq_21_01}
\end{equation}
so that the mass parameters $m_{LL}^2$, $m_{RR}^2$, $m_{LR}^2$ can also be
expressed via the measurable observables $m_{\tilde{\tau}_{1,2}}^2$:
\begin{eqnarray}
m_{LL}^2&=&\frac{m^2_{\tilde{\tau}_1}+m^2_{\tilde{\tau}_2}}{2}
-\frac{m^2_{\tilde{\tau}_2} - m^2_{\tilde{\tau}_1}}{2}
\cos(2 \theta_{\tilde{\tau}}),\la{eq_m_ll}\\
m_{RR}^2&=&\frac{m^2_{\tilde{\tau}_1}+m^2_{\tilde{\tau}_2}}{2}
+\frac{m^2_{\tilde{\tau}_2} - m^2_{\tilde{\tau}_1}}{2}
\cos(2 \theta_{\tilde{\tau}}),\la{eq_m_rr}\\
m_{LR}^2&=&\frac{1}{2}(m_{\tilde{\tau}_1}^2-m^2_{\tilde{\tau}_2})
\sin(2 \theta_{\tilde{\tau}}).\la{eq_m_lr}
\end{eqnarray}
One sees from (\ref{eq_m_ll}) and (\ref{eq_m_rr}) that one can distinguish
the two cases
\begin{enumerate}
\item if $m^2_{LL}>m^2_{RR}$ then
$\cos(2\theta_{\tilde{\tau}})=-\frac{1}{\sqrt{1+\xi^2}}<0$
\item if $m^2_{LL}<m^2_{RR}$ then
$\cos(2\theta_{\tilde{\tau}})=+\frac{1}{\sqrt{1+\xi^2}}>0$
\end{enumerate}
The cross section for $\tilde{\tau}_i\tilde{\tau}_i$ production
depends on the mixing angle $\cos 2 \theta_{\tilde{\tau}}$ (\cite{Wien}
and references therein)
and $\cos\theta_{\tilde{\tau}}$ can be accurately determined, with a
two--fold ambiguity, via the cross section for $\tilde{\tau}_i$
production with polarized beams or 
via a polarization asymmetry
$A_{Pol}=(\sigma_L-\sigma_R)/(\sigma_L+\sigma_R)$.

In the following we discuss the $\tau$ 
polarization from $\tilde{\tau}_i$ decays $P_{\tilde{\tau}_i\to \tau}$.
The tau polarization
$P_{\tau}$ can be derived from the energy distribution of the
decay products of the tau lepton.

\vspace{-.2cm}
\subsection{\bm ${\tau}$ \um polarization from \bm $\tilde{\tau}_i$ \um decays}
\vspace{-.2cm}
In this section we study the polarization $P_{\tau}$
from the decays
\begin{eqnarray}
&&\tilde{\tau}_{1}\to \tilde{\chi}^0_i \tau, \label{eq_22_1a}
\quad\mbox{and}\quad
\tilde{\tau}_{2}\to \tilde{\chi}^0_i \tau,\quad i=1,\ldots,4
\end{eqnarray}
taking into account a
general neutralino mixing in the MSSM \cite{Boos}.

The tau polarization for (\ref{eq_22_1a}) is given by 
(using the narrow width approximation) \ci{Noji}:
\begin{eqnarray}
P_{\tilde{\tau}_1\to \tau}&=&
\frac{(a_{1i}^R)^2-(a_{1i}^L)^2}{(a_{1i}^R)^2+(a_{1i}^L)^2},\label{eq_22_2a}
\quad\mbox{and}\quad 
P_{\tilde{\tau}_2\to \tau}=
\frac{(a_{2i}^R)^2-(a_{2i}^L)^2}{(a_{2i}^R)^2+(a_{2i}^L)^2}.
\end{eqnarray}
with
\begin{eqnarray}
a_{1i}^{L,R}&=&\cos\theta_{\tilde{\tau}} a_{Li}^{L,R}
+\sin\theta_{\tilde{\tau}} a_{Ri}^{L,R}, \label{eq_22_2c}
\quad\mbox{and}\quad
a_{2i}^{L,R}=-\sin\theta_{\tilde{\tau}} a_{Li}^{L,R}
+\cos\theta_{\tilde{\tau}} a_{Ri}^{L,R}, 
\end{eqnarray}
where the coefficients $a_{ij}^{k}$ are defined by the Lagrangian
\begin{equation}
{\cal L}=\sum_{{i=1,2 \atop j=1,\ldots,4}}
\tilde{\tau}_i \bar{\tau}(P_L a^R_{ij}+P_R a^L_{ij})\tilde{\chi}^0_j.
\label{eq_22_3}
\end{equation}
in the neutralino basis $(\tilde{B},\tilde{W}_3,\tilde{H}_1,\tilde{H}_2)$:
\begin{eqnarray}
a_{Lj}^R&=&- \frac{g}{\sqrt{2}} \frac{m_{\tau}}{m_W \cos\beta} N_{j3},\quad
\quad\quad
a_{Rj}^L=a_{Lj}^R \label{eq_22_4a}\\
a_{Lj}^L&=&+ \frac{g}{\sqrt{2}} [N_{j2}+N_{j1} \tan\theta_W],\quad
a_{Rj}^R=- \frac{2 g}{\sqrt{2}} N_{j1} \tan\theta_W
\label{eq_22_4d}.
\end{eqnarray}
Taking into account a general mixing in the neutralino sector
we derive the $\tau$ polarization: 
\begin{eqnarray}
P_{\tilde{\tau}_1\to \tau}&=&\frac{(4-x_W^2)-(4+x_W^2-2
y_h^2)\cos{2\theta_{\tilde{\tau}}}+2(2+x_W)y_h\sin{2\theta_{\tilde{\tau}}}}
{(4+x_W^2+2y_h^2)-(4-x_W^2)
\cos{2\theta_{\tilde{\tau}}}+2(2-x_W)y_h\sin{2\theta_{\tilde{\tau}}}}
\label{eq_22_5a}.
\end{eqnarray}
And for the case  $m_{LL}^2>m_{RR}^2$ the formula takes the form:
\begin{eqnarray}
\phantom{P_{\tilde{\tau}_1\to \tau} }&=&\frac{(4-x_W^2)\sqrt{1+\xi^2}+(4+x_W^2-2
y_h^2)-2(2+x_W)y_h\xi}{(4+x_W^2+2 y_h^2)\sqrt{1+\xi^2}
+(4-x_W^2)-2(2-x_W)y_h\xi} \label{equ_22_5a},
\end{eqnarray}
where the mixing angle $\theta_{\tilde{\tau}}$ i.e. $\xi$
is given by (\ref{eq_21_01}) and $P_{\tilde{\tau}_2\to \tau}$ can be obtained 
from  eq.~(\ref{eq_22_5a}) by changing the sign of 
$\cos{2\theta_{\tilde{\tau}}}$ and $\sin{2\theta_{\tilde{\tau}}}$. 

The coefficient $x_W$ contains the complete contribution 
from the gaugino components, 
$y_h$ is a combination of the factors of the Yukawa coupling
and the complete contribution from the higgsino 
components $x_h$.
\begin{eqnarray}
x_W&=&\frac{\tan\theta_W N_{11}+N_{12}}{\tan\theta_W N_{11}}
\label{eq_22_6b}\\
y_h&=&\frac{\frac{1}{\cos\beta}\frac{m_{\tau}}{m_W}~N_{13}}{\tan\theta_{W}~N_{11}}
=:\frac{1}{\cos\beta}\frac{m_{\tau}}{m_W}x_h. \label{eq_22_6e}
\end{eqnarray}
One sees from eq.~(\ref{eq_22_5a})
that the transformation between the
two cases $m_{LL}^2>m_{RR}^2$, 
whose hierarchy is motivated by  the MSSM as far as 
no addional D--terms have to be included, and
 $m_{LL}^2<m_{RR}^2$ lead only
to an exchange of $P_{\tilde{\tau}_1\to \tau}\leftrightarrow 
P_{\tilde{\tau}_2\to \tau}$. 

\subsubsection{\bm $\tan\beta$ \um dependence of the \bm $\tau$ \um 
polarization}
\vspace{-.1cm}
With the coefficients $x_W$ and $x_h$ in
eqn.~(\ref{eq_22_5a})--(\ref{equ_22_5a}) the {\it complete}
$\tan\beta$ dependence from the neutralino sector has been separated
and the coefficient $y_h$ shows the interplay between the Yukawa
coupling and the $\tilde{\chi}^0_i$ higgsino admixture.
All components of (\ref{eq_22_6b}), (\ref{eq_22_6e})
are given explicitly as function of the fundamental
MSSM parameters in \cite{CKMZ}
and in an approximation, 
which is valid for high values of
$\tan\beta$ in \cite{Boos}.\\[-.5em]

\noindent{\it One can summarize} that the dependence on $\tan\beta$ of 
$P_{\tilde{\tau}_{1,2}\to \tau}$ is given in a three-fold way:\\[-1.8em]
\begin{enumerate}
\item by the $\tan\beta$ dependence of the mixing angle 
$\theta_{\tilde{\tau}}$ (\ref{eq_21_01}) and 
the off-diagonal term in (\ref{eq_21_002});
\item by 
the coefficients $x_W$ and $x_h$, which corresponds to the 
$\tan\beta$ dependence of the neutralino sector,
\item by the coefficient $y_h$ which corresponds to the
$\tilde{\tau}$ Yukawa coupling   
$\frac{1}{\cos\beta}\frac{m_{\tau}}{m_W}$.
\end{enumerate}
One sees clearly
from (\ref{eq_22_6e}) that for a given 
mixing angle $\theta_{\tau}$ or
$\xi$ only the coefficient 
$y_h$ contains a strong dependence on
$\tan\beta$ of $P_{\tilde{\tau}_1\to \tau}$ from the Yukawa coupling.
Obviously a sufficient large higgsino admixture part $x_h$ has to be
for that.

The $\tan\beta$ dependence of $x_W$ and $x_h$ is weak since
the neutralino mass eigenvalues $m_i^2$ as well as the
components of the eigenvectors $N_{ij}$ become in the high $\tan\beta$
approximation $f(1+const\times \frac{1}{\tan \beta})$.
Therefore it is a good approximation to
take all neutralino mixing contributions in the high $\tan\beta$
approximation. 

\vspace{-.2cm}
\subsection{Limiting cases: extrema for the mixing angle}
\begin{figure}[t]
\setlength{\unitlength}{1cm}
\begin{center}
\begin{minipage}{7cm}
\hspace*{2.5cm}
\begin{picture}(7,9)
\put(-.2,0){\includegraphics{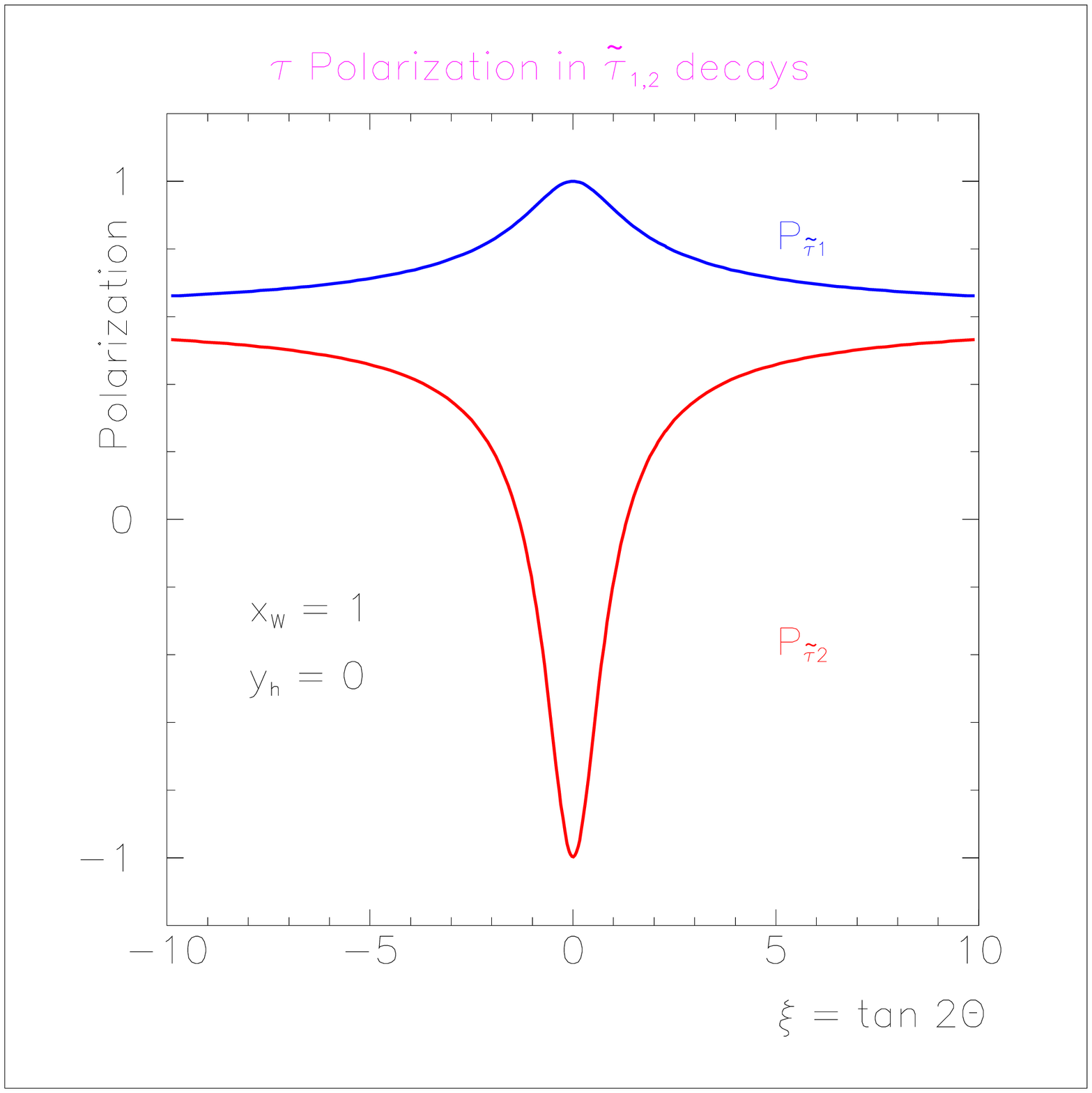}}
\end{picture}\par\vspace{.4cm}\hspace{-1cm}
\end{minipage}
\hspace*{-.5cm}
\begin{minipage}{7cm}
\begin{picture}(7,9)
\put(-.2,0){\includegraphics{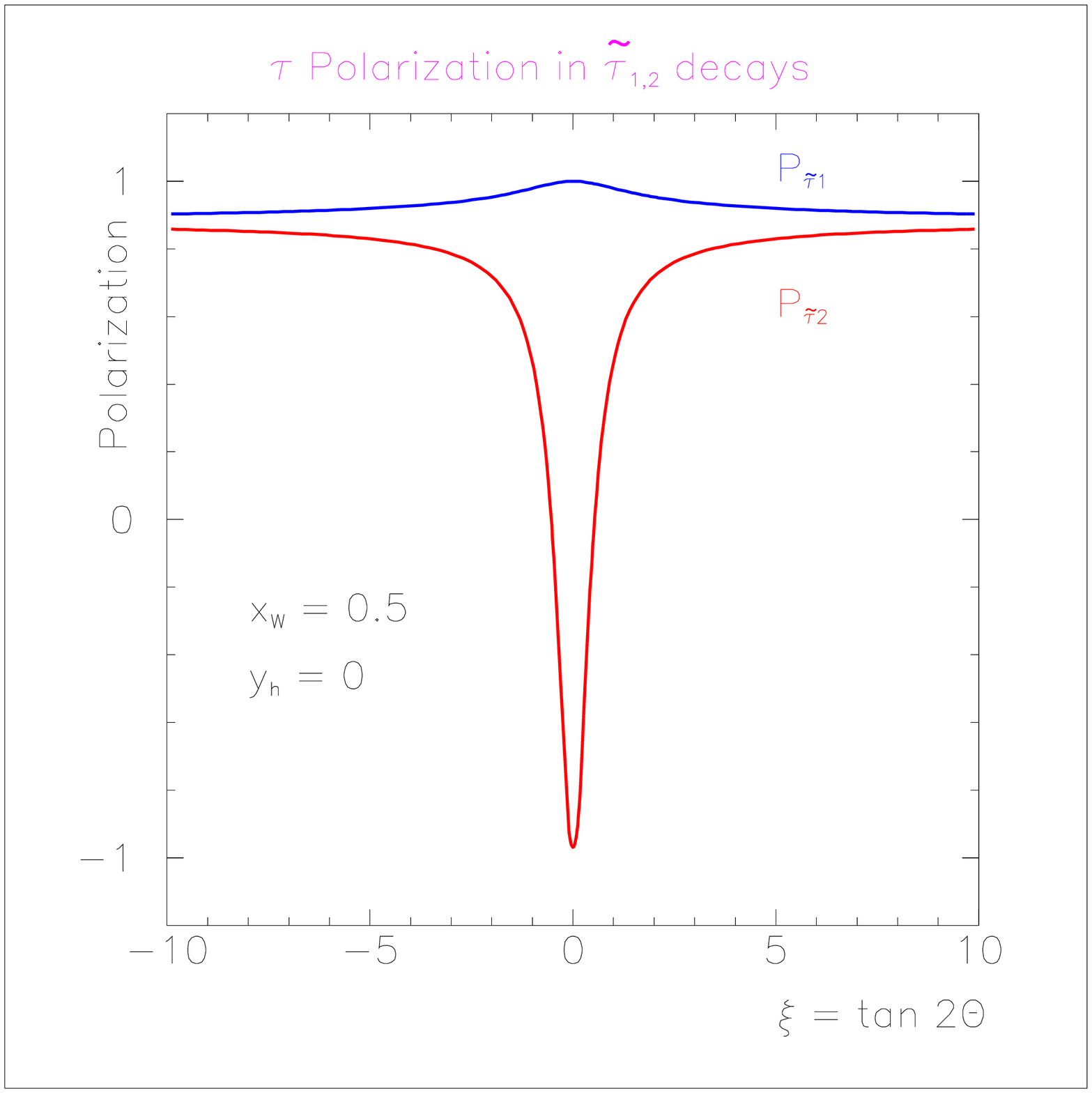}}
\end{picture}\par\vspace{.4cm}\hspace{-1cm}
\end{minipage}
\end{center}\vspace*{-4cm}
  \caption{\label{fig_bino} The dependence of the  
tau polarizations $P_{\tilde{\tau}_1\to\tau}$ and
$P_{\tilde{\tau}_2\to\tau}$  from the mixing angles $\xi$ for a pure bino case 
(left) with the neutralino mixing variables $x_W=1$, $y_h=0$ 
and for a pure gaugino case (right) with $x_W=0.5$, $y_h=0$. }%
\end{figure}

\vspace{-.2cm}
Assuming no constraints for the $\tau$
mass parameters one can vary the mixing angle $\theta_{\tilde{\tau}}$
from 0 to $\pi$, i.e. $\xi$ from $0\to \pm \infty$, 
and we get for these extrema the following expressions:
\begin{eqnarray}
&&P_{\tilde{\tau}_1\to\tau}^{\xi \to \pm \infty}
=\frac{4-x_{W}^2+ 2~(2+x_W)~y_h}
{4+x_{W}^2-2~(+ 2-y_h - x_W)~y_h }\la{eq_22_7a} \\
&&P_{\tilde{\tau}_1\to\tau}^{\xi \to 0}
=\frac{4-y_h^2}{4+y_h^2}\la{eq_22_7b}\\
&&P_{\tilde{\tau}_2\to\tau}^{\xi \to \pm \infty}
=\frac{4-x_{W}^2- 2~(2+x_W)~y_h}
{4+x_{W}^2+2~(+ 2-y_h- x_W)~y_h } \la{eq_22_7c}\\
&&P_{\tilde{\tau}_2\to\tau}^{\xi \to 0}
=\frac{-x_W^2+y_h^2}{x_W^2+y_h^2} \la{eq_22_7d}
\end{eqnarray}
The formulae for the tau polarization even simplifies for specific
neutralino mixing cases. We give the corresponding expressions in
Table~\ref{tab_mix} for the processes $\tilde{\tau}_1\to\tau\tilde{\chi}^0_1$
and $\tilde{\tau}_2\to\tau\tilde{\chi}^0_1$. The conventions are chosen so that
in a no--mixing case, i.e. 
$\tilde{\tau}_1\to \tilde{\tau}_R<\tilde{\tau}_2\to \tilde{\tau}_L$,
and when only a pure U(1) gauge coupling interacts between $\ell
\tilde{\ell}\tilde{\chi}^0$, the polarization $P_{\tilde{\tau}_1^-\to
\tau_R^-}=+1$ and $P_{\tilde{\tau}_2^-\to \tau_L^-}=-1$.  Studying the
polarization of the antiparticle $\tau^+$ one has to take into account
that the SUSY partner of $\tau^+_{L,R}$ is $\tilde{\tau}^+_{R,L}$, so
that $P_{\tilde{\tau}_R^+\to \tau_L^+}=-1$ and $P_{\tilde{\tau}_L^+\to
\tau_R^+}=+1$. We show in Fig.~\ref{fig_bino} representative examples for 
illustration. 

\begin{table}
\begin{center}
\begin{tabular}{|cc|c|c|c||c|c|c|}
\hline
& $\tilde{\chi}^0_1$ & $P_{\tilde{\tau}_1\to \tau}$ &
$P_{\tilde{\tau}_1\to \tau}^{\xi \to \pm \infty}$ &
$P_{\tilde{\tau}_1\to \tau}^{\xi \to 0}$ &
$P_{\tilde{\tau}_2\to \tau}$ &
$P_{\tilde{\tau}_2\to \tau}^{\xi \to \pm \infty}$ &
$P_{\tilde{\tau}_2\to \tau}^{\xi \to 0}$ \\ \hline 
&&&&&&&\\
a) & $\approx N_{11}$ & $\frac{3\sqrt{1+\xi^2}+5}{5\sqrt{1+\xi^2}+3}$ & 
$\frac{3}{5}$ & $+1$ & $\frac{3\sqrt{1+\xi^2}-5}{5\sqrt{1+\xi^2}-3}$ & 
$\frac{3}{5}$ & $-1$ \\ \hline
&&&&&&&\\
b) & $\approx c_{11} N_{11}$ & 
$\frac{(4-x_{W}^2)\sqrt{1+\xi^2} +(4+x_{W}^2)}
{(4+x_{W}^2)\sqrt{1+\xi^2} +(4-x_{W}^2)}$ & 
$\frac{4-x_W^2}{4+x_W^2}$ & $ +1$ & 
$\frac{(4-x_{W}^2)\sqrt{1+\xi^2} -(4+x_{W}^2)}
{(4+x_{W}^2)\sqrt{1+\xi^2} -(4-x_{W}^2)}$ & $\frac{4-x_W^2}{4+x_W^2}$
& $-1$ \\
 & $+c_{12} N_{12}$ & &&&&&\\ \hline 
&&&&&&&\\
c) & $\approx N_{13}$ & $\frac{-1}{\sqrt{1+\xi^2}}$  
& 0 & $-1$ & $\frac{+1}{\sqrt{1+\xi^2}}$ & 0 & +1 \\ \hline
\end{tabular}
\caption{$\tau$ polarization for extrema for the mixing angle and
specific neutralino mixing.
\label{tab_mix}}
\end{center}\vspace*{-1cm}
\end{table}
The corresponding plot for case a), Table~\ref{tab_mix}, 
is given in Fig.~\ref{fig_bino}a. The asymptotic limit 
$\frac{3}{5}$ for $\xi \to \pm \infty$ can clearly be seen.

The corresponding plot for case b), Table~\ref{tab_mix}, is given in
Fig.~\ref{fig_bino}b. The
asymptotic limit depends now on the gaugino parameters and since the
wino fraction $x_{W}^2$ is less than 1 the polarization
$P_{\tilde{\tau}_1\to\tau}$ gets closer to 1 and will always be higher
than the asyptotic value $\frac{3}{5}$ of the pure bino case a).

In the pure higgsino case c), Table~\ref{tab_mix},
the no--stau--mixing case shows a helicity flipping behaviour,
$P_{\tilde{\tau}_1\to\tau}\to -1$ and $P_{\tilde{\tau}_2\to\tau}\to +1$,
which is the typical feature of the Yukawa couplings.

\begin{figure}
\setlength{\unitlength}{1cm}
\begin{center}
\begin{minipage}{7cm}
\hspace*{2.5cm}
\begin{picture}(7,9)
\put(-.2,0){\includegraphics{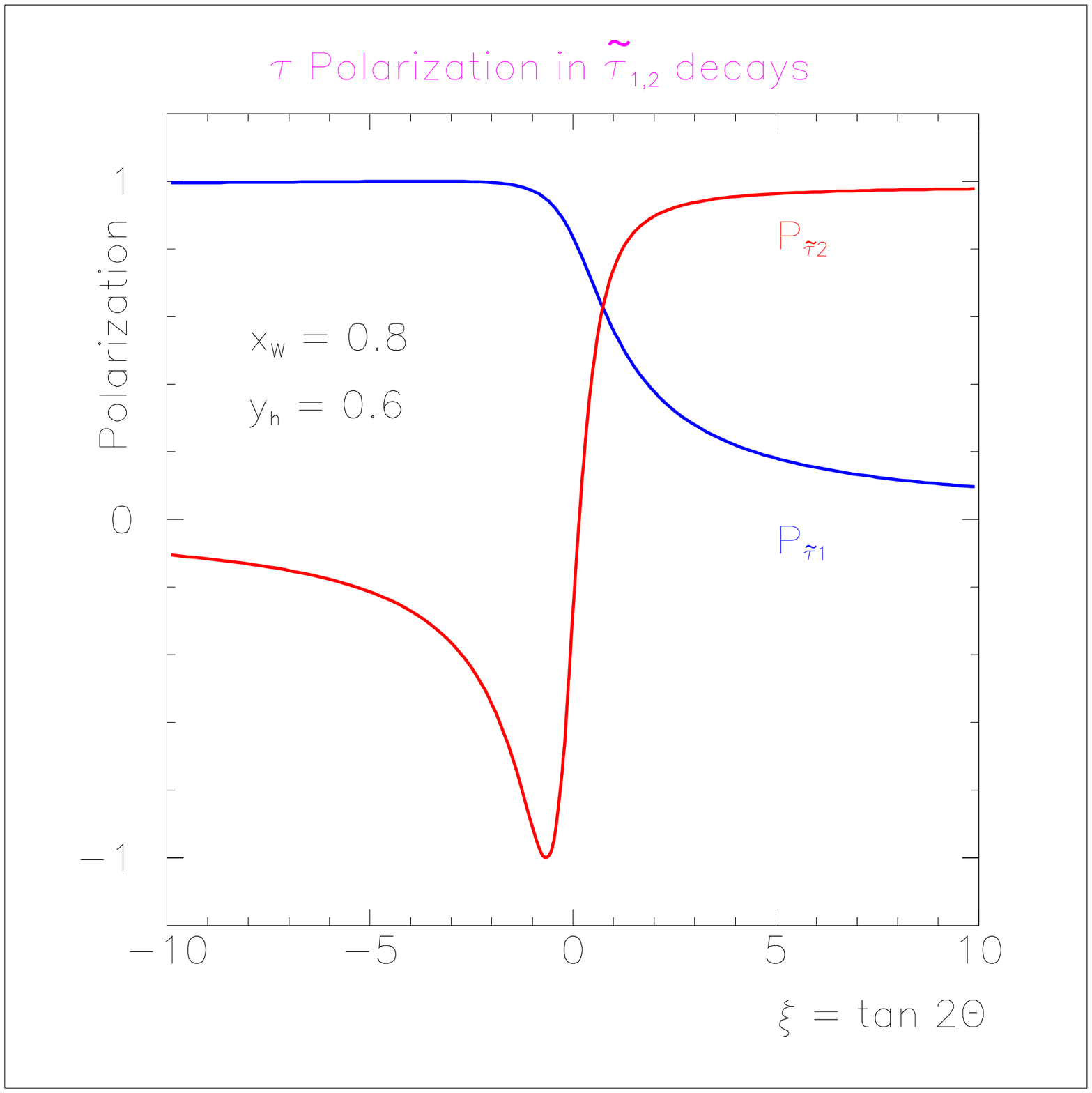}}
\end{picture}\par\vspace{.4cm}\hspace{-1cm}
\end{minipage}
\hspace*{-.5cm}
\begin{minipage}{7cm}
\begin{picture}(7,9)
\put(-.2,0){\includegraphics{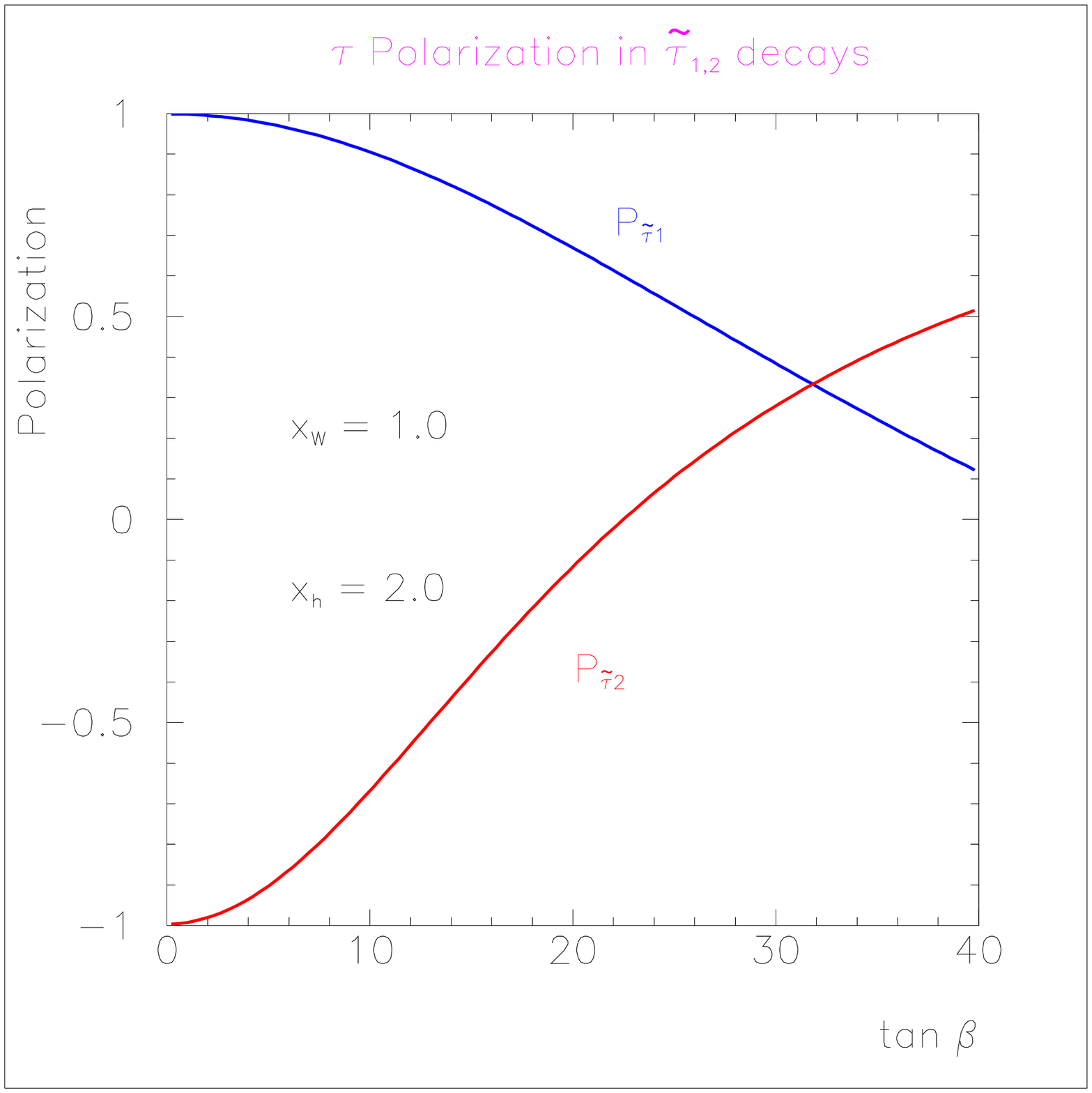}}
\end{picture}\par\vspace{.4cm}\hspace{-1cm}
\end{minipage}
\end{center}\vspace*{-4cm}
  \caption{\label{fig-mix} The dependence of the tau polarizations
$P_{\tilde{\tau}_1\to\tau}$ and $P_{\tilde{\tau}_2\to\tau}$ from the
mixing angles $\xi$ (left) for a mixed case with the neutralino
mixing variables $x_W=0.8$, $y_h=0.6$ and as function of $\tan\beta$
(right) where the mixing angle is chosen to be small $\xi\to 0$. In
this case $\tilde{\tau}_1\to \tilde{\tau}_R$, $\tilde{\tau}_2\to
\tilde{\tau}_L$.}%
\end{figure}

To give a
feeling even for the neutralino mixing effects 
we plot in Fig.~\ref{fig-mix}a the
polarizations $P_{\tilde{\tau}_1\to\tau}$, $P_{\tilde{\tau}_2\to\tau}$ 
for the mixed case $x_W=0.8$ and $y_h=0.6$.  The polarizations
even interchange for a specific mixing angle.

So if the fraction of the higgsino mixing part is small compared to the 
gaugino mixing one expects $P_{\tilde{\tau}_1\to \tau}$ 
close to unity and $P_{\tilde{\tau}_2\to\tau}$ 
variable in a large range, eqn.~(\ref{eq_22_7a})--(\ref{eq_22_7d}). 
This is the case for all the SPS scenarios 
\ci{SPS} as shown in the Table~\ref{tab-sps}. 

In order to demonstrate the $\tan\beta$ 
dependence coming from the interplay between 
the Yukawa coupling and the higgsino 
admixture of the LSP 
we show in Fig.~\ref{fig-mix}b
the polarizations $P_{\tilde{\tau}_1\to\tau}$ and
$P_{\tilde{\tau}_2\to\tau}$ 
for the case with a rather large higgsino admixture and set
$\xi\to 0$, i.e. $\tilde{\tau}_1\to
\tilde{\tau}_R$ and $\tilde{\tau}_2\to \tilde{\tau}_L$.

\begin{table}[h]
  \begin{center}
\begin{tabular}{|c|c|cc|cc|}
\hline
Parameter Point&$\tan\beta$ 
&\multicolumn{2}{c|}{$\tau$ Polarization}&\multicolumn{2}{c|}{slopes}\\
\cline{3-6}
          & &~~$P_{\tilde{\tau}_1\to\tau}$ & ~~$P_{\tilde{\tau}_2\to\tau}$ & $d(P_{\tilde{\tau}_1\to\tau})/d(\tan\beta)$ & 
$d(P_{\tilde{\tau}_2\to\tau})/d(\tan\beta)$ \\  \hline
SPS 1a & 10   & 98.1 \%   &    -50 \% & $-0.3\%$ & 5.0\% \\
SPS 1b & 30   & 97.0 \%   &    -40 \% & $-0.1\%$ & 1.6\% \\
SPS 3  & 10   & 99.2 \%   &    -80 \% & $-0.1\%$ & 2.4\% \\
SPS 4  & 50   & 99.6 \%   &    -62 \% & +0.1\%   & $-2.0\%$   \\
SPS 5  & 5    & 97.8 \%   &    -60 \% & $-0.6\%$ & 7.0\% \\
SPS 6  & 10   & 99.0 \%   &    -65 \% & $-0.2\%$ & 4.0\% \\
\hline
\end{tabular}\vspace{-.5cm}
   \end{center}%
   \caption{\label{tab-sps}%
Tau polarization and $\tan\beta$ slopes for the SPS scenarios \cite{SPS}.
The program ISAJET 7.58 \cite{Isa} has been used
for the parameter evaluation. The SPS 2 focuspoint point scenario
does not have a stable slope.}
\vspace*{-.3cm}
\end{table}
In order to get an impression in which ranges of the parameter space
the tau polarizations from $\tilde{\tau}_{1,2}$ decays may have 
large variations,
we show in Fig.~\ref{martin_cont} the polarizations
$P_{\tilde{\tau}_{1,2}\to\tau}$ as functions of  $M_2$ and
$\mu$. The other relevant
parameters have been chosen to $\tan\beta=40$, $M_L=300$~GeV,
$M_E=150$~GeV, $A_{\tau}=-254.2$~GeV.  Both plots show that one could
get high values for the polarization for a large region of the
$M_2-\mu$ parameter space.  
In particular in the higgsino--like
region with $\mu<M_2$ the polarization $P_{\tilde{\tau}_1\to \tau}$  
is very variable.

%% file: Strategy_fin.tex
\section{MSSM parameter determination for high \bm $\tan\beta$ \um }
\vspace{-.2cm}
\subsection{Parameter determination in the \bm $\tilde{\tau}_{1,2}$ \um sector}
\vspace{-.2cm}
We assume that one can measure at a high ${\cal L}$ LC
the masses with rather high accuracy of about $\%$ level via mass
threshold scans.

We study the light system $e^+ e^-\to \tilde{\tau}_1 \tilde{\tau}_1$ 
and determine the mixing angle $\cos\theta_{\tilde{\tau}}$ via polarized 
cross sections $\sigma(e^+_{L,R}e^-_{R,L}\to\tilde{\tau}_1 \tilde{\tau}_1)\sim
\cos(2 \theta_{\tilde{\tau}})$ or via
the asymmetry $A_{pol}=(\sigma_L-\sigma_R)/(\sigma_L+\sigma_R)$. 
If one wants to derive
$\theta_{\tilde{\tau}}$ unambigously one would also need
$\sigma(e^+e^-\to\tilde{\tau}_1 \tilde{\tau}_2)\sim
\sin(2 \theta_{\tilde{\tau}})$, which is more tricky because of the 
difficult reconstruction of the $\rho$, $\pi$ decay products 
from the $\tilde{\tau}_2$ decays.

\begin{figure}
\setlength{\unitlength}{1cm}
\begin{center}
\begin{picture}(7,9)
\put(-1,0){\includegraphics{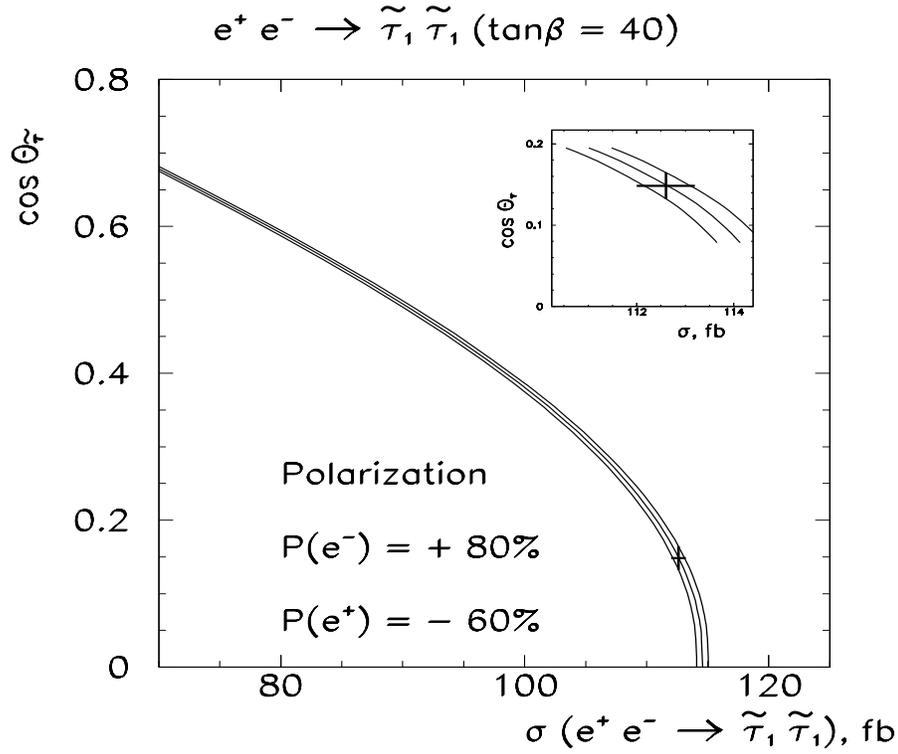}}
\end{picture}\par\vspace{.4cm}\hspace{-1cm}
\end{center}
\vspace*{-2.5cm}
  \caption{\label{sigpol} Mixing angle $\cos\theta_{\tilde{\tau}}=0.15\pm 0.01$
via measurement of the polarized cross section
$\sigma(e^+e^-\to\tilde{\tau}_1\tilde{\tau}_1)$ at $\sqrt{s}=500$~GeV and
$P(e^-)=+80\%$, $P(e^+)=-60\%$ and on the assumption that
1 standard deviation as statistical error was taken into account.
}%
\end{figure}

We determine the mixing angle via measuring the production
$e^-e^+\to\tilde{\tau}_1^-\tilde{\tau}_1^+$ in the configuration
$(RL)$, $P(e^-)=+80\%$, $P(e^+)=-60\%$, since in this case the worse
background from $W^+W^-$ is strongly suppressed and in contrary the
signal is enhanced.  From the rates $\sigma_{RL}$ we could derive the
mixing angle rather accurately, $\cos{\theta_{\tau}}=0.15\pm 0.01$,
Fig.\ref{sigpol}\footnote[1]{Generically the cross section is a
quadratic polynom in $\cos(2 \theta_{\tilde{\tau}})$. With a suitably
high degree of beam polarization, however, the quadratic terms are
suppressed and only one solution survives.}, if we measure
$\sigma_{RL}(\tilde{\tau}_1^-\tilde{\tau}_1^+)=112$~fb at
$\sqrt{s}=500$~GeV and assume 
that a statistical error of $\pm 1 \sigma$ was taken into account. 
In Table~\ref{tab_rates} we list the corresponding
cross 
sections $\sigma(\tilde{\tau}_{i}\tilde{\tau}_{j})$ 
for $\sqrt{s}=500$~GeV and $\sqrt{s}=800$~GeV. 

In principle one can alternatively
also derive the mixing angle via measuring the 
polarization asymmetry $A_{pol}$ of the production process. 
Taking rates with left polarized electrons
leads, however, to an enhancement of the  
strong $WW$ background and a tricky analysis
would be needed to get the wanted  experimental information.

\begin{table}[b]
\begin{center}
\begin{tabular}{|c|c||c|c|c|}
\hline
 & \multicolumn{1}{c|}{$\sqrt{s}=500$~GeV} & 
\multicolumn{3}{|c|}{$\sqrt{s}=800$~GeV}\\ 
$(P(e^-),P(e^+))$ & $\sigma(e^-e^+\to\tilde{\tau}_1\tilde{\tau}_1)$ & 
$\sigma(e^-e^+\to\tilde{\tau}_1\tilde{\tau}_1)$ & 
$\sigma(e^-e^+\to\tilde{\tau}_1\tilde{\tau}_2)$ & 
$\sigma(e^-e^+\to\tilde{\tau}_2\tilde{\tau}_2)$\\ \hline
unpolarized &48.6 fb&29.7 fb&0.2 fb& 12.0 fb\\
$(-0.8,0)$ &25.6 fb&15.9 fb&0.3 fb& 18.3 fb \\
$(+0.8,0)$ &71.6 fb&43.5 fb &0.2 fb & 5.7 fb \\
$(-0.8,0.6)$ & 31.6 fb &19.8 fb&0.4 fb&28.8 fb \\
$(+0.8,-0.6)$ &112.1 fb &68.1 fb &0.3 fb& 6.7 fb\\ \hline
\end{tabular}
\caption{Cross sections 
for the reference scenario with polarized beams. The pairs
$\tilde{\tau}_1\tilde{\tau}_2$, although kinematically accessible
at $\sqrt{s}=500$~GeV, lead to rates less than $0.1$~fb. 
\label{tab_rates}}\vspace{-1cm}
\end{center}
\end{table}
\begin{figure}
\setlength{\unitlength}{1cm}
\begin{center}
\begin{picture}(15,10)
\put(0,0){\includegraphics{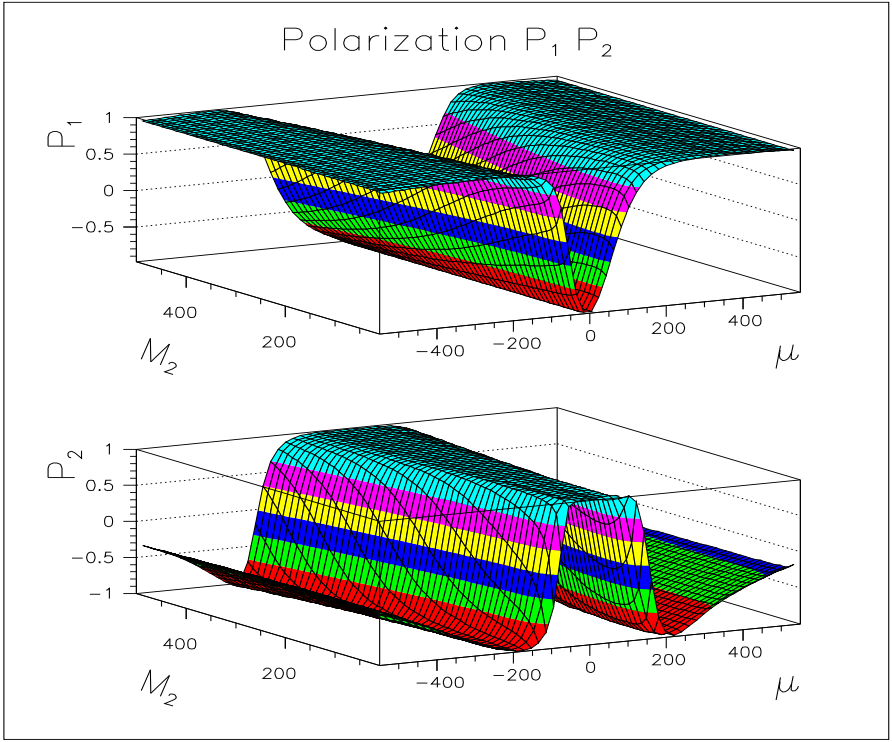}}
\end{picture}
\end{center}\vspace*{-3cm}
  \caption{\label{martin_cont} Contourplots in the $M_2$-$\mu$ plane for 
$P_{\tilde{\tau}_1\to\tau}$ and $P_{\tilde{\tau}_2\to\tau}$. }%
\end{figure}

\subsection{Measurement of the \bm $\tau$ \um polarization}
We have computed the cross sections and disctributions
for the 4 final state particle processes with all spin correlations
taken into account:
\begin{equation}
e^+ e^- \to \tilde{\tau}^+_1 \tilde{\tau}^-_1 \quad\mbox{with}\quad
\tilde{\tau}_1^-\to \tau^-_{L,R} \tilde{\chi}^0_1\quad\mbox{and}\quad 
\tau_{L,R}^- \to \nu_\tau \pi^-
\end{equation}
In order to 
measure the polarization of the $\tau$'s one has to study their decays 
into $\pi$ or $\rho$--mesons and to fit kinematically
the energy distributions of the decay 
products \cite{Noji}.
The decays of polarized $\tau$ to $\pi$- and $\rho$-mesons
are implemented into CompHEP and are cross checked with TAUOLA
\cite{TAUOLA}.

The calculations are performed by means of
the CompHEPV41 \cite{comphep}  with implemented MSSM and 
mSUGRA models. 
In order to estimate an accuracy of polarization measurements 
we have generated a number of unweighted events which correspond to the
production cross section with the corresponding decay branching fractions
and the overall registration efficiency of signal events:
\begin{equation}
N = \sigma(e^+ e^- \rightarrow \tilde{\tau}^-_{1} \tilde{\tau}^+_{1})
* Br (\tilde{\tau}^-_{1} \rightarrow  \tau \tilde{\chi}^0_1)
* Br ( \tau \rightarrow \nu_{\tau}\pi^-)* {\cal L}* \mbox{eff}
\label{event}
\end{equation}
 Based on the results of MC analysis given in \cite{Noji}
we assumed 30\% for the efficiency and we assume $500 fb^{-1}$ 
for a luminosity. So in total about 3300 umweighted events
have been simulated. 

The pion energy distribution is shown in the Fig.~\ref{tau_pol} left.
The distribution is described by the formular presented in a slightly
different form as that one given in \cite{Noji}.
\begin{eqnarray}
&&\frac{1}{\sigma}\frac{d\sigma}{dy_{\pi}}(e^+ e^- \rightarrow
\tilde{\tau}^-_{1} \tilde{\tau}^+_{1}
\rightarrow \tau \tilde{\chi}^0_1
\tilde{\tau}^+_{1}
\rightarrow \nu_\tau \pi^- \tilde{\chi}^0_1
\tilde{\tau}^+_1) \nonumber \\
&&= \frac{1}{x_{max}-x_{min}} \label{eq_fit}
\begin{cases}
(1-P_{\tilde{\tau}_1\to\tau})log\frac{x_{max}}{x_{min}} +
2P_{\tilde{\tau}_1\to\tau}y_{\pi}(\frac{1}{x_{min}}-\frac{1}{x_{max}}),
&\text{$0<y_{\pi}<x_{min}$}\\
(1-P_{\tilde{\tau}_1\to\tau})log\frac{x_{max}}{y_{\pi}} +
2P_{\tilde{\tau}_1\to\tau}(1-\frac{y_{\pi}}{x_{max}}),
&\text{$x_{min}<y_{\pi}$}\\
\end{cases}
\end{eqnarray}  
where $y_{\pi}=\frac{2E_{\pi}^{CMS}}{\sqrt{s}}$,
$x_{min}=\frac{2E_{\tau}^{min}}{\sqrt{s}}$, 
$x_{max}=\frac{2E_{\tau}^{max}}{\sqrt{s}}$, 
$E_{\tau}^{max,min}=\frac{E_{\tau}^* \pm p_{\tau} \beta_{\tilde{\tau}}}
{\sqrt{1-\beta_{\tilde{\tau}}^2}}$,
$\beta_{\tilde{\tau}}=\sqrt{1-\frac{4M_{\tilde{\tau}}^2}{s}}$,
$E_{\tau}^* = \frac{M_{\tilde{\tau}}^2 - M^2_{\tilde{\chi}^0_1} +
M_{\tau}^2}{2 M_{\tilde{\tau}}}$,
$p_{\tau} = \sqrt{E_{\tau}^{*2}- M_{\tau}^2}$.\\
  
The fit of the energy distribution by the above formula (\ref{eq_fit}) 
gives for the tau 
polarization $P_{\tilde{\tau}_1\to\tau}$ about $57\%\pm 3\%$, 
compared to the theoretical value of polarization for the
reference point of $56\%$. 

The presented example is meant as an illustration only.
Measurements of 
the $\tau$ lepton decay mode to  $\rho$ with its subsequent decays
may help to get even a better accuracy. 
\subsection{Determination of \bm $\tan\beta$ \um and the \bm $A_{\tau}$ \um
parameter}
\begin{figure}
\setlength{\unitlength}{1cm}
\begin{center}
\begin{minipage}{7.5cm}
\begin{picture}(7,9)
\put(-.2,0){\includegraphics{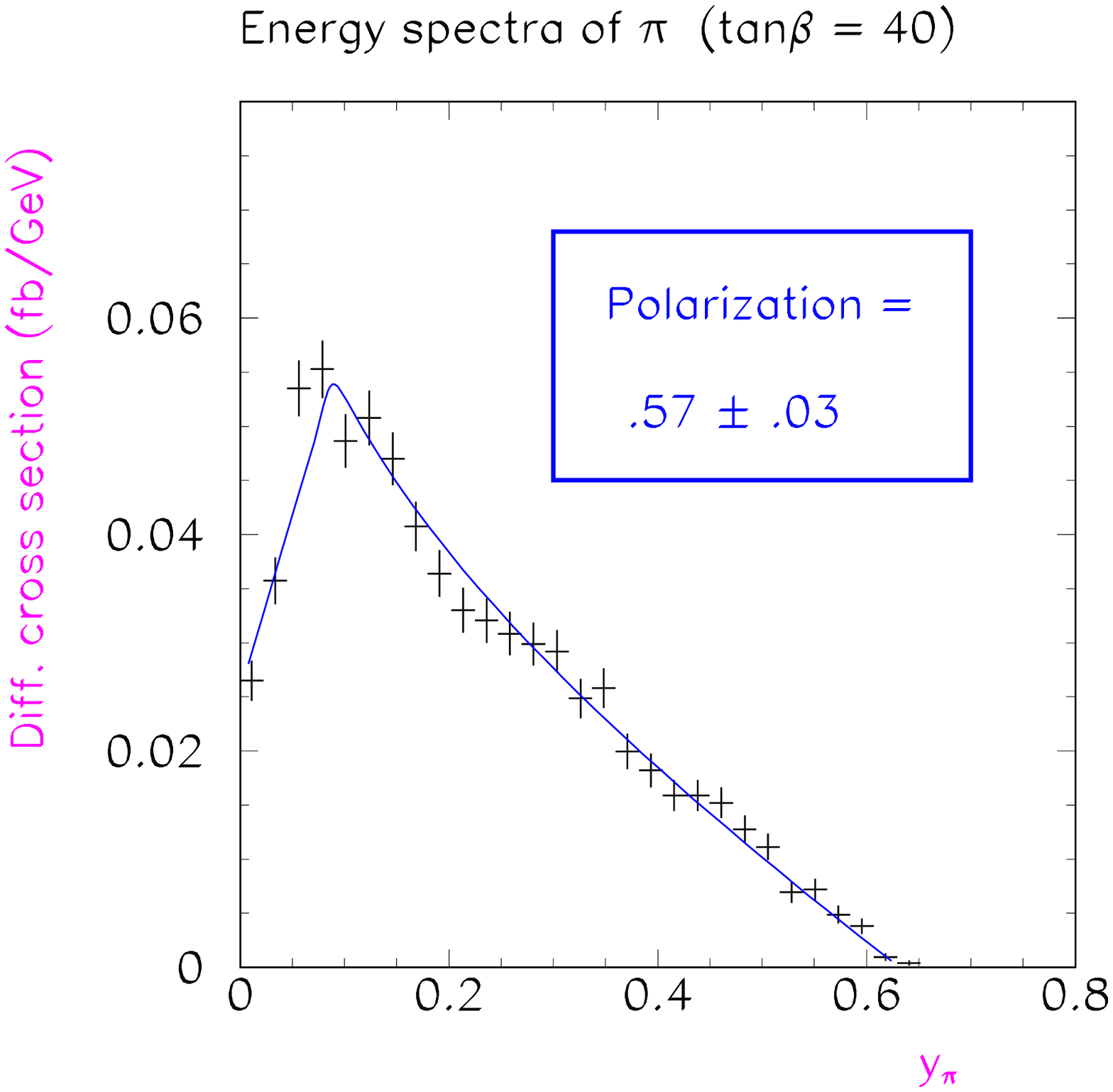}}
\end{picture}
\end{minipage}\vspace{-6.5cm}
\hspace*{2cm}
\begin{minipage}{7cm}
\begin{picture}(7,9)
\put(-.2,0){\includegraphics{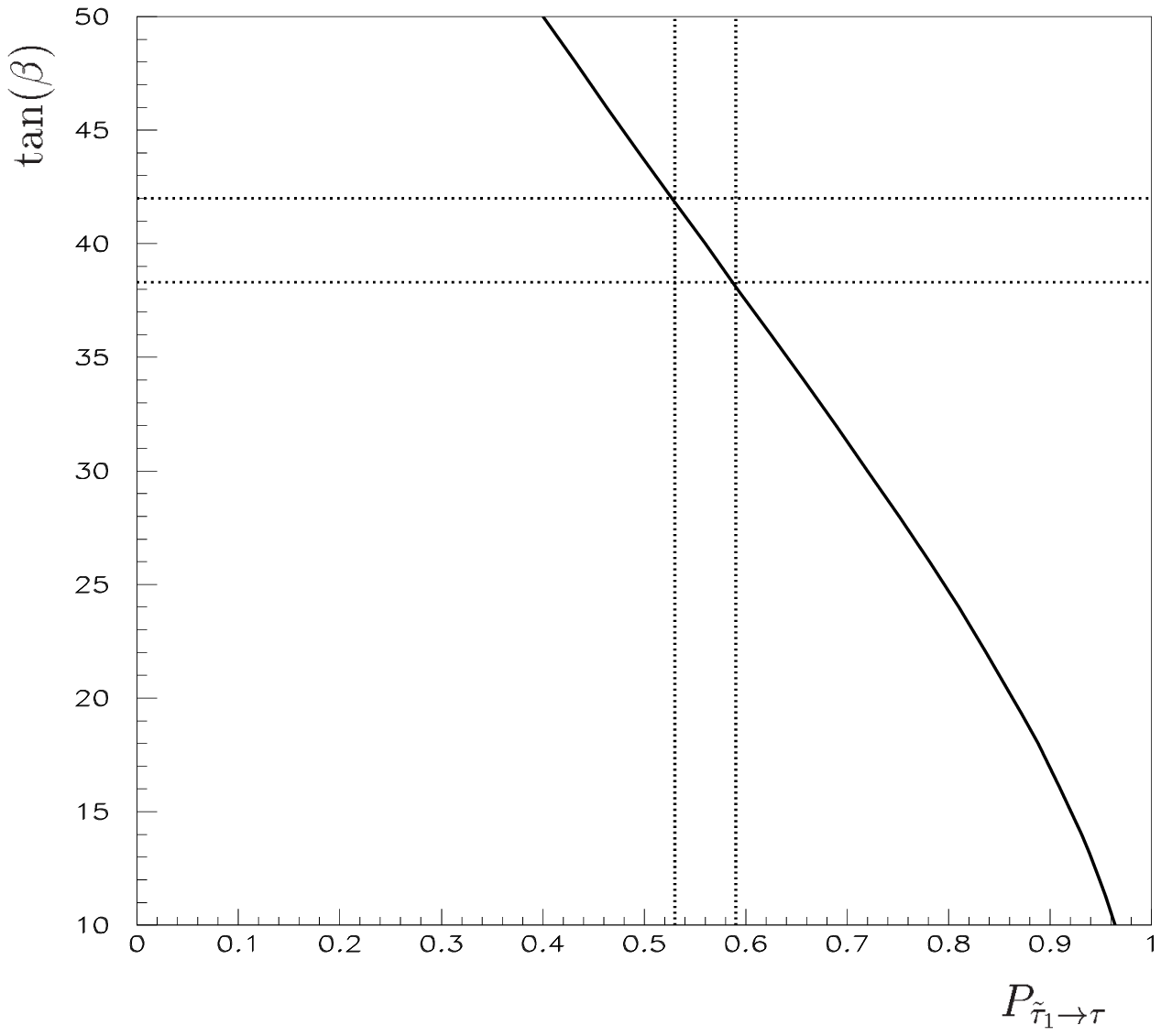}}
\end{picture}
\end{minipage}\vspace{-6.5cm}
  \caption{\label{tau_pol}
Left: The pion energy histogram for the 
unweighted
events and the 1~$\sigma$ fit by the distribution formula (\ref{eq_fit}). 
The total event number is 3300 events which corresponds to an 
integrated luminosity of ${\cal L}=500$~fb$^{-1}$. For beam polarization 
we choose 
$P_{e^-}= +80\%$, $P_{e^+}= -60\%$.
 Right: The determination of $\tan\beta=40\pm 2$ as function of the measured
polarization $P_{\tilde{\tau}_1\to\tau}=57\pm 3\%$. 
}%
\end{center}
\end{figure}

In the last section we have shown that 
$P_{\tilde{\tau}_1\to\tau}$ could be measured within about
$5\%$ accuracy.
We derive the value of $\tan\beta$ via inversion of (\ref{eq_22_5a})
for the calculated values of $x_W$ and $x_h$ and the measured 
mixing angle $\cos\theta_{\tilde{\tau}}$, see Fig.~\ref{tau_pol} right, and 
we determine in our scenario the high $\tan\beta$ with an error of about
$5\%$:
\begin{equation}
\tan\beta=40\pm 2  
\label{eq_tb}
\end{equation}

In the case that also the heavier mass of $\tilde{\tau}_2$ can be 
determined in the experiment,
we can even determine the parameter $A_{\tau}$ without assuming anything 
about the SUSY breaking scheme and GUT relations:
\begin{equation}
A_{\tau}=\frac{1}{m_{\tau}}(\frac{1}{2}
(m_{\tilde{\tau}_1}^2-m_{\tilde{\tau}_2}^2)\sin(2\theta_{\tau})+m_{\tau}
\mu \tan\beta).
\label{eq_atau}
\end{equation}
With an error of about $5\%$ in $\tan\beta$ and of about
$5\%$ in $m_{\tilde{\tau}_2}$ one could derive 
$A_{\tau}$ with an accuracy about $8\%$.
For the application of this method on the $\tilde{b}$ and $\tilde{t}$ sector
and a detailed simulation study see \cite{Boos}.

%% file: Conclusions_fin.tex
\section{Conclusions}
\vspace{-.2cm}
Tau polarization from stau decays might be an important variable
for the determination of the fundamental MSSM parameters.
It serves as a `bridge' between chargino/neutralino and stau sectors.

The stau mixing angles can be precisely determined via polarized rates
and in  case that the decay neutralino has a suitable higgsino admixture,
the study of $P_{\tilde{\tau}_{1}\to \tau}$
leads to an accurate determination of $\tan\beta$ even in the case of high
$\tan\beta$. 
For these procedure it is enough to measure only the light system:
$m_{\tilde{\tau}_1}$ and $\sigma(e^+e^-\to\tilde{\tau}_1\tilde{\tau}_1)$.
In case that also $m_{\tilde{\tau}_2}$ can be measured,
even the parameter $A_{\tau}$ can be derived without an assumption about 
the underlying SUSY breaking scheme.

For a given example we explored $P_{\tilde{\tau}_{1}\to \tau}$ and 
showed that one can determine simultanously $A_{\tau}$ and $\tan\beta$, 
e.g. in the case of high $\tan\beta$,  within $5\%$ accuracy.

A systematic simulation of the process as well as the application 
of this method on the $\tilde{b}$ and $\tilde{t}$ sector
will be presented in a following paper \cite{Boos}.